# Local compression properties of double-stranded DNA based on a dynamic simulation

Xiaoling Lei[1*], Wenpeng Qi[1] and Haiping Fang[1]

[1] *Division of interfacial water and Laboratory of Physical Biology, Shanghai Institute of Applied Physics, Chinese Academy of Sciences, P.O. Box 800-204, Shanghai 201800, China.*

The local mechanical properties of DNA are believed to play an important role in their biological functions and DNA-based nanomechanical devices. Using a simple sphere-tip compression system, the local radial mechanical properties of DNA are systematically studied by changing the tip size. The compression simulation results for the 16 nm diameter sphere tip are well consistent with the experimental results. With the diameter of the tip decreasing, the radial compressive elastic properties under external loads become sensitive to the tip size and the local DNA conformation. There appears a suddenly force break in the compression-force curve when the sphere size is less than or equal to 12 nm diameter. The analysis of the hydrogen bonds and base stacking interaction shows there is a local unwinding process occurs. During the local unwinding process, first the hydrogen bonds between complement base pairs are broken. With the compression aggregating, the local backbones in the compression center are unwound from the double helix conformation to a kind of parallel conformation. This local unwinding behavior deducing by external loads is helpful to understand the biological process, and important to DNA-based nanomechanical devices.

*leixiaoling@sinap.ac.cn

## I. INTRODUCTION

The mechanical properties of DNA are essential in many biological functions such as DNA packing, replication and transcription, are not fully understood [1, 2]. During these processes, the binding proteins induce local deformations of the DNA double helix [3]. Insight into the local mechanical properties of DNA under external loads will helpful to understanding the relative biological functions. Furthermore the local mechanical properties of DNA play a key role in DNA-based molecular nanowires, arrays, and objects [4-6].

The mechanical properties of double-stranded DNA (ds-DNA) molecules have been extensively studied with the new experimental tools such as the atomic force microscopy [7, 8], the small-angle x-ray scattering [9] and the magnetic tweezer [10]. Most experiments have been focused on the stretching, bending and twisting properties of DNA. Zhou and Wang have studied the response of a DNA molecule in the radial direction using vibrating scanning polarization force microscopy (VSPFM), which is believed to be helpful in understanding the local elastic deformation property of ds-DNA [11, 12]. Recent results of the experiments and the theories indicate that the local mechanical properties of ds-DNA (less than one persistence length) still remain controversial. Results of some experiments indicate that the local bending stiffness of DNA can be as much as 3–5 folds lower than anticipated from the Wormlike Chain model [13, 14], while others support that the local bending stiffness is standing on the anticipated from the Wormlike Chain model [15].The study of the bending property indicates that the local mechanical properties of ds-DNA still remain intriguing and controversial, and need new approaches to understanding.

In this letter we present a systematic study of the radial compression properties of ds-DNA using a sphere-tip compression system that we recently proposed [16]. The simulation is based on a discrete elastic model of ds-DNA at the base pair level together with a langevin dynamic. A model ds-DNA segment of 91 base pairs is compressed by a sphere tip whose diameter changes from 16 nm to 2 nm. The compression-force curves show that the simulation results for the 16 nm diameter sphere tip are well consistent with the experimental results. When we decrease the diameter of the sphere tip, the radial elastic rigidity under external loads gradually exhibits sensitive to the tip size and the local ds-DNA conformation. We find a suddenly force break in the compression-force curve when we decrease the sphere tip size less than or equal to 12 nm diameter. The changing of the hydrogen bonds and base stacking interaction suggests that the local unwinding process occursthrough the coupling of local hydrogen bonding breaking and neighbor base pairs relaxing helix. This results shed light upon the understanding of the biological function of ds-DNA and have important implications for the possible mechanisms of ds-DNA.

## II. THE DISCRETE ds-DNA MOLECULE MODEL

A ds-DNA molecule is a linear polymer of deoxyribonucleotides: two single-stranded DNA (ss-DNA) molecules wind each other right-handedly around a common central axis to form a ds-DNA molecule in which hydrogen bonds between complementary bases connect the ds-DNA molecule tightly. Our discrete model is based on the model proposed by Zhou, Yang and Ouyang [17]. We consider the sugar-phosphate backbones of a ds-DNA molecule as two extensible wormlike chains with the same elasticity modulus and same bending rigidity. Each chain is modeled by a trace of beads connected by semirigid springs. Each bead has the mass of a deoxyribonucleotides with a position of the mass center of the deoxyribonucleotides, which is307 a.m.u.. The springs are assumed massless, and the lengths are determined by the distance between neighboring beads. The structure parameters used in the ds-DNA molecule model are as follows: diameter of 2.0 nm, screw-pitch of 3.4 nm, and each turn containing 10 bps as shown in Fig.1.

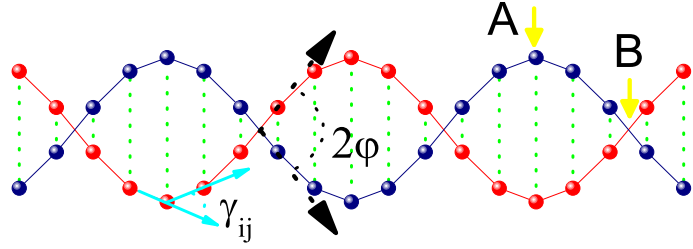

FIG. 1. Schematic view of the discrete model. The sugar-phosphate backbone of the ds-DNA molecule is represented as a wormlike chain and each chain is modeled by a trace of beads connected by springs (red and navy lines). Two ss-DNA molecules are connected by the hydrogen bonds (green dotted line) to form a double-stranded conformation. $\varphi$ is the folding angle of the sugar-phosphate backbones which is half the rotational angle between two backbones at the same length. $\gamma_{ij}$ is the bending angle between the neighboring springs of the same chain. The yellow arrows marked by A and B denote the local conformation position where the sphere tip is placed in the simulation.

The total energy includes the backbone bending energy $E_b$, the backbone stretching energy $E_e$, the hydrogen bond energy between the complementary bases $E_v$, the base stacking energy between neighboring base-pairs $E_{bp}$, and the excluded-volume energy $V_m$. The total energy of a ds-DNA molecule without external force is defined as:

$$E = E_b + E_e + V_m + E_{bp} + E_v. \qquad (1)$$

For a continuous chain, the backbone bending energy can be described by $(k_B T l_p/2) \int_0^l (\partial t_j/\partial s)^2 ds$, where $t_j(s)$ ($j = 1\ and\ 2$) is the unit tangent vector at arclengths along the $j$th backbone [17], $l_p$ is the persistent length of the backbone [18]. In this study, the discrete form of the backbone bending energy can be written as

$$E_b = \frac{k_B T l_p}{2} \sum_{j=1,2} \sum_{i}^{N-2} \gamma_{ij}^2. \qquad (2)$$

Where N is the number of beads (nucleotides) of each backbone, $i$ is the number of springs and $j$ is the number of chains. $\gamma_{ij}$ is the bending angle between the neighboring springs of the same chain as shown in Fig1. In this paper, we set $lp$ = 12.0 nm (A detailed description on the effect of $lp$ can be found in reference [19]).

A simple harmonic potential is used to describe the backbone stretching energy of the backbone extension for the discrete model:

$$E_e = \sum_{j=1,2} \sum_{i}^{N-1} \frac{1}{2} k_e (l_{ij} - l_0)^2, \qquad (3)$$

where $l_{ij}$ is the distance between $i$th and $(i+1)$th beads of $j$th backbone, $l_0$ is the equilibrium distance between neighboring beads along the backbone; it is measured as 0.705 nm in our simulations. $ke$ = 18.72 eV is relative to the elasticity modulus of ss-DNA molecules[20].

One of the most popular potential models to describe the hydrogen bonds interaction is the Morse potential [21]. The total hydrogen bonding energy for the discrete model is

$$V_m = \sum_{i=1}^{N} D_0 (e^{-\alpha(y_i - y_0)} - 1)^2, \qquad (4)$$

where $y_i$ is the length between the base pairs, and $y_0$ is the initial length which equals the diameter of a ds-DNA. $D_0$ is the depth of the Morse potential well, which may depend on the type of base pair (A-T or G-C), and $\alpha$ describes the width of the well. If we only consider the average sense and take them as constants, the depth of the Morse potential well $D_0$ and the width of the well $\alpha$ would be $D0$

= 0.22 eV and $\alpha$ = 3.5 nm$^{-1}$[17].

The base stacking interaction between two adjacent base pairs has been described via the Lennard-Jones-type potential [17]:

$$\rho_i = \begin{cases} \epsilon\left[\left(\frac{\cos\varphi_0}{\cos\varphi_i}\right)^{12} - 2\left(\frac{\cos\varphi_0}{\cos\varphi_i}\right)^{6}\right] \varphi_i \geq 0, \\ \epsilon[\cos\varphi_0^{12} - 2\cos\varphi_0^{6}]\varphi_i < 0, \end{cases} \quad (5)$$

where $\varphi_i$ is the folding angle of the sugar-phosphate backbones which is half the rotational angle between two backbones at the same length as depicted in Fig. 1, $\varphi_0$ is the equilibrium folding angle ( ~62$^0$), and $\epsilon = 0.36$ eV is the base stacking intensity [22]. The total base stacking energy is $E_{bp} = \sum_1^{N-1} \rho_i$.

To avoid each backbone overlaps with itself or the other one, the excluded-volume effects have been considered using a simple rigid sphere potential:

$$E_v = \sum_{i=5}^{N-5} \sum_{k=-4}^{k=5} \zeta \left(\frac{r_0}{r_{ik}}\right)^9. \quad (6)$$

Here, $\zeta$ = 0.0063 eV. $r_{ik}$ is the distance between $(k+i)th$ bead of one chain and $ith$ bead of the other chain. We take $r_0$ = 1.0 nm, which equals to the radius of ds-DNA.

We apply the langevin dynamics to simulate the dynamic process [23]:

$$m\dot{v}(t) = F(t) - \lambda v(t) + \Gamma(t), \quad (7)$$

where $F = -\nabla E$, here $E$ is the energy as shown in eq.(1), $\lambda = 0.0001$ is the friction constant. $\Gamma(t)$ is the randomforce, which is produced from the Gaussian distribution with a standard variance at the simulation temperature $\langle \Gamma_i(t)\Gamma_i(t')\rangle = 6\lambda k_B T \delta(t-t')$[24], $t$ is the simulation time, and $\delta(t-t')$ is the Dirac delta function. Every bead is subjected to a random force at each integration time step. The components of the random force are independently generated by setting $\Gamma_{ij} = \tau \Phi \sqrt{2m\lambda k_B T/dt}$, where $j$ denotes the three components of random force in the x, y, and z direction, $\Phi$ is a random value taken from a standard Gaussian distribution (zero mean and unit variance),and *dt*is the integration time step. The velocity and position of each bead are updated at each dynamics time step by using a half-step "leap-frog" scheme [25]. The time step *dt*in our simulation is 0.02 ps. The speed of the compression is fixed in all of the simulations, which is $v_c = 10^6$ um/s.

## III. DNA MOLECULE UNDER RADIAL COMPRESSION

A sphere-tip system we developed [16] is used in this paper as shown in Fig. 2 (a). A ds-DNA segment of 91 base pairs placed on a rigid solid surface was compressed by a sphere tip graduall falling with a constant velocity just above the ds-DNA molecule. A repel interaction between the ds-DNA molecule and the sphere tip is

$$U_{tip} = \sum_{i=1}^{N} \xi \left(\frac{d_0}{d_i}\right)^9. \tag{8}$$

Here, $\xi = 0.0063\ eV$. $d_i$ is the distance between the surfaceof the sphere tip and the center of each bead. $d_0 = 0.34\ nm$ is the equilibrium distance which is about equal to the van der Waals diameter of a carbon atom.

We present the systematic study of the external mechanical compression upon local ds-DNA when the tip size can be compared with ds-DNA diameter. By changing the sphere diameter from 16 to 2 nm, the ds-DNA gradually show different compression properties from the experimental result [11, 12] and our previous work [16]. Figure 2(b) shows the simulation results of different compression size together with the experimental data. It is clearly that the compression property is sensitive to the sphere tip size and local DNA conformation. The compression force is reduced from the maxium1.6 nN to 0.1 nN with the tip size decreasing from 16 nm to 2 nm sphere diameter; therefore, the growth of force with tip sizes agrees with simple physical intuition for smaller tip deducing smaller compression force.

Each solid and open squares of same color in Fig. 2(b) correspond to the compression vs. force curves for the same size tip placing in two extreme ds-DNA positions A and B as shown in Fig. 1. The divergence between two positions is growing from ignorable for 20 nm diameter tip to significant for 2 nm diameter tip. The greatest divergence between two positions occurs at around 0.8 nm ds-DNA height for different tip size, and the divergence increases from about 30 pN for 12 nm diameter tip to 200 pN for 4 nm diameter tip. Clearly, we can figure out the local conformation of ds-DNA from the radial compression elastic property of small tip size. The result is instructive for the experiment related to local DNA property and DNA-based nanotechnology [26].

In contrast to the experimental results, a suppressing force break gradually appears with the compression tip size decreasing. After this sudden force break, the compression force slowly increase until another force suddenly decrease again. The suddenly force break means the ds-DNA suddenly become more soft, which is out of our image. The analyzing of the folding angle in axis direction and the length between the base pairs in radial direction will benefit us to understanding the force break. Figure 3 shows the length between the base pairs at varying heights of DNA for different tip size located at different positions. The four compression measures of different tip size exhibit very similar results all showing a length oscillations between the base pairs. Greater compression intensifies the oscillation range. Furthermore the difference for two positions of same sphere size can be ignored inthe length changing between the base pairs. The length changing between the base pairs represents the trend of hydrogen bonding interactions which is particularly important in terms of ds-DNA molecules' stability and determining the specificity of DNA-protein recognition [27, 28]. Obvious length oscillation between the base pairs means the hydrogen bond has been weakened and broken. Local conformation transition occurred through the breaking of the hydrogen bonds, and this transition minimizes the inner strain aggregated during the compression. The behavior demonstrates that base stacking, not base pairing, is of primary importance for the radial elastic properties of ds-DNA under external loads. This similar behavior has also been found in other studies such as the excited-state dynamics in A▪T DNA oligonucleotides [29, 30].

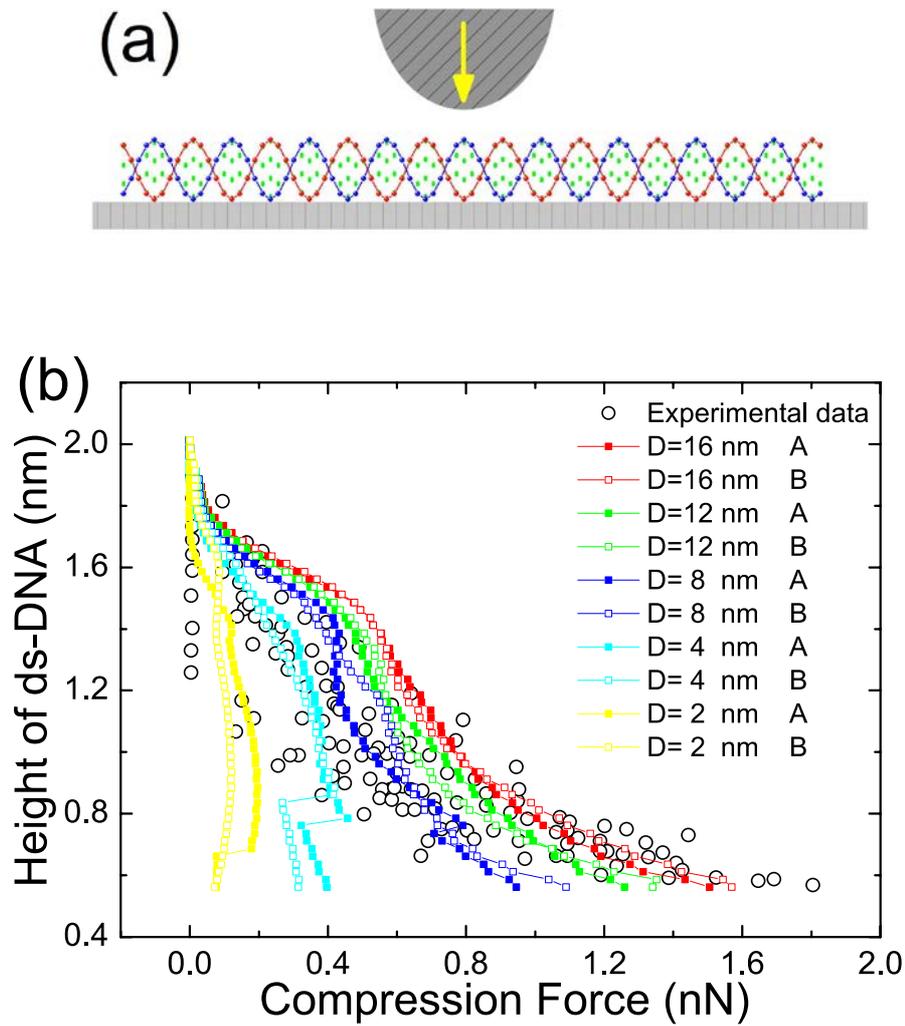

FIG. 2. (a) A schematic view for compressing the ds-DNA molecule model system. The sphere tip (gray semi-circle) compresses the ds-DNA molecule lying on the rigid surface (gray rectangle) along the vertical direction (yellow arrow). (b) Comparison of the simulation results with the experimental result in refs. [11, 12]. The color lines denote the simulation results compressed by different sphere tip size, the solid and open squares of same color denote the sphere tip placed in A and B positions as shown in Fig. 1. The black open circles denote the experimental results of 20 ds-DNA segments.

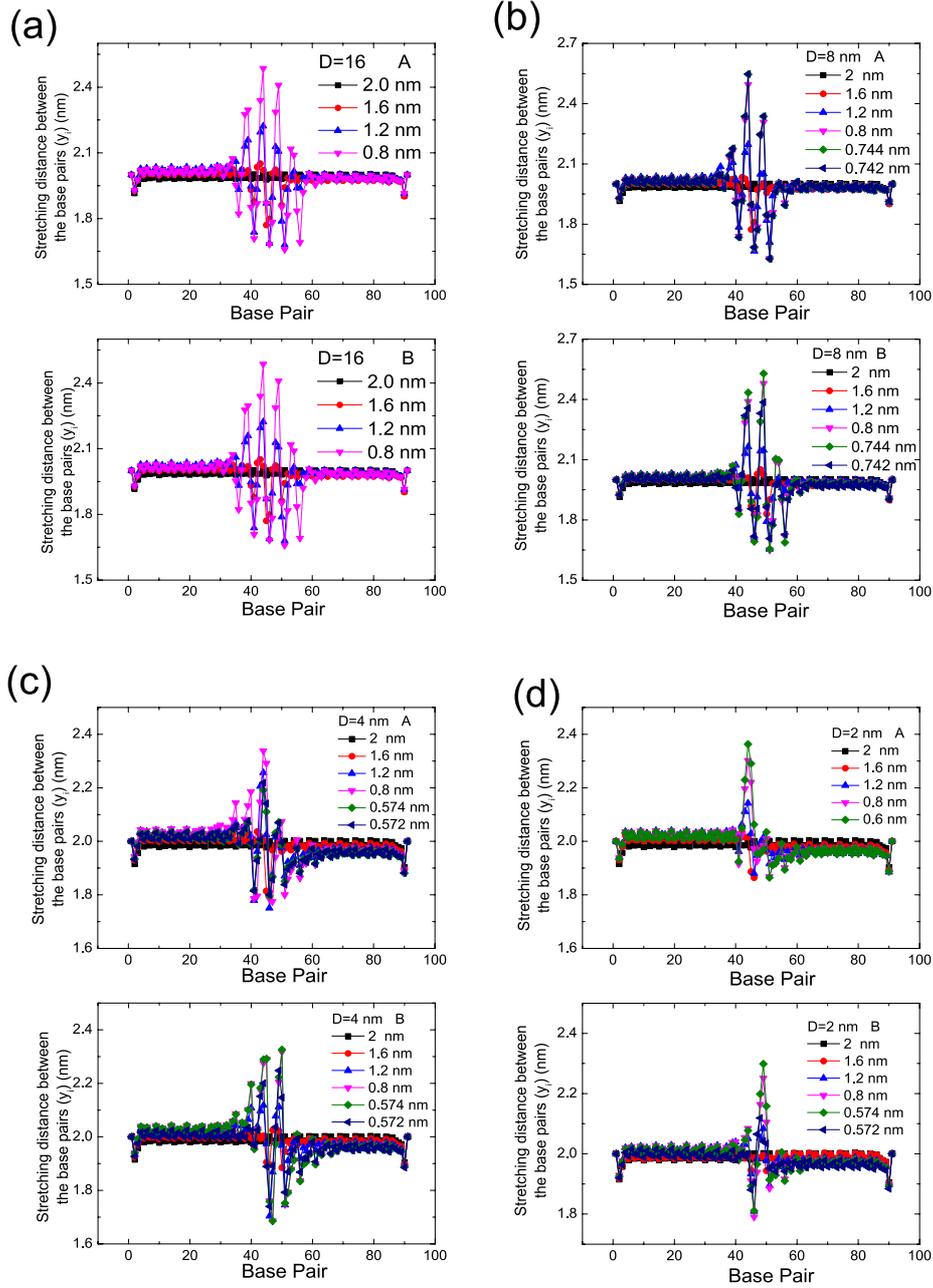

FIG. 3. The length of complementary base pairing for the ds-DNA molecules at different heights during the compression processes. Different color lines denote the changing at different compression heights. (a), (b), (c) and (d) correspond to the sphere size of 16 nm, 8 nm, 4 nm and 2 nm diameter respectively. The upper and lower curves of each image correspond to the sphere tip placed at position A and B.

Base stacking interaction is one of the important sources of ds-DNA molecule's double-helix stability and biological function [8, 31]. The decrease in folding angles indicates the weakening of base stacking interaction in the ds-DNA molecule's axis direction. Figure 4 shows the results of folding angles at varying heights of DNA for different tip size located at different positions. The results for diameter 16 nm sphere size are exactly similar with experimental results [16], and the folding angles change slightly about 5% percent during the compression process. When the sphere-tip size reducing, a clearly difference appears in the folding angle changing. Most folding angles change slightly except a few folding angles in the compression center suddenly decrease from the initial value (about $62^0$ in our model) to about $20^0$ to $30^0$ just when the compression force reaches to the force break. Combination of Fig. 2(b) with Fig. 4 suggests that the folding angle decreasing just occurs when the force break is shown in the compression property. We conclude that local backbones are unwound from the double helix conformation to a kind of parallel conformation through the coupling of local hydrogen bonding breaking and neighbor base pairs relaxing helix. This is the reason that the force break appears in the compression process which due to ds-DNA suddenly become soft. The unwind behaviors are very important in many biological processes such as DNA replication, gene regulation and DNA-proteininteraction [2, 32]. The locally unwind behavior of ds-DNA under external loads shown in our result is helpful to understand the biological process, furthermore is important to DNA-based nanomechanical devices [33, 34].

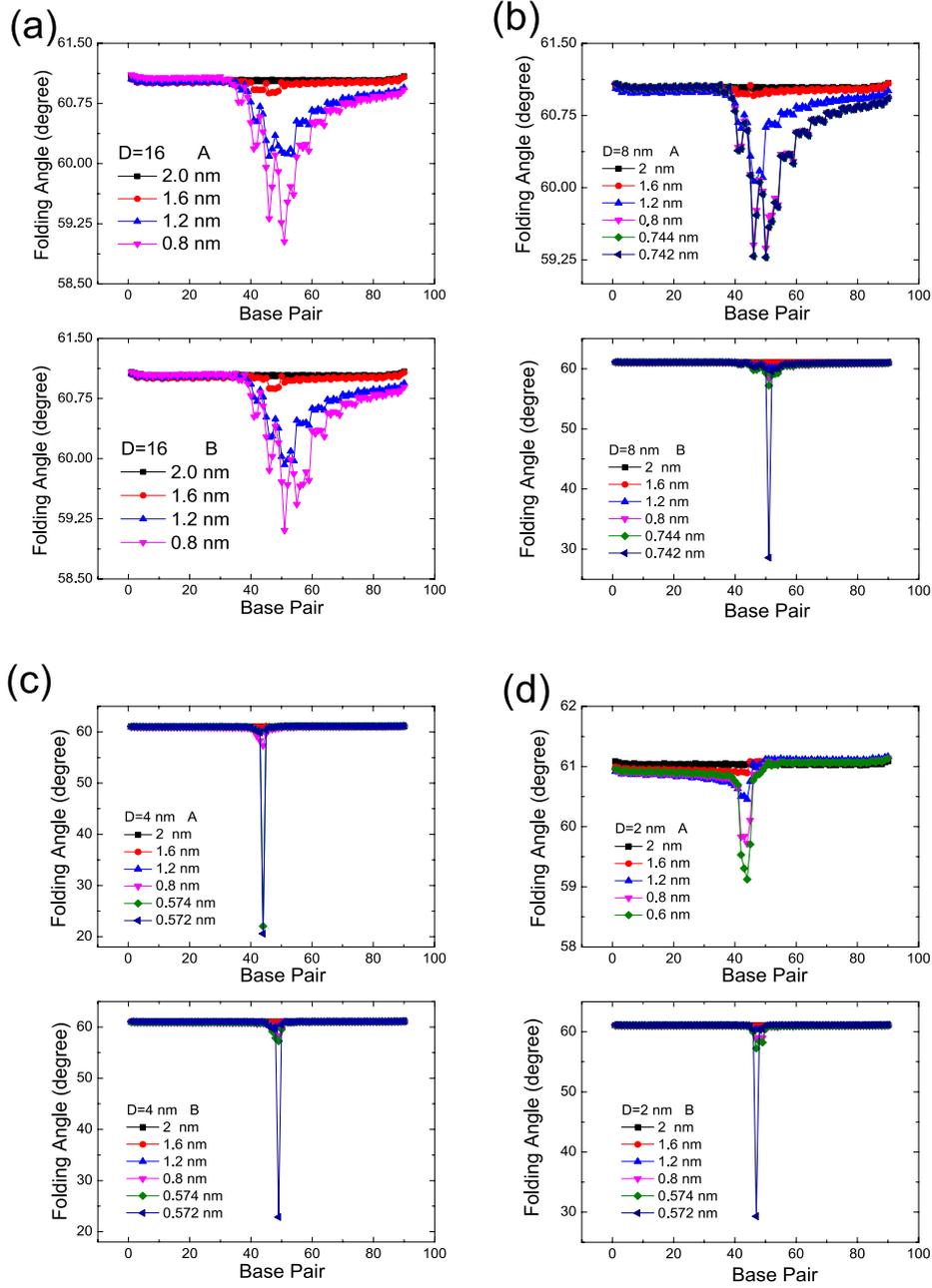

FIG. 4. The folding angle between two neighboring base pairs for the ds-DNA molecules at different heights during the compression processes. Different color lines denote the changing at different compression heights. (a), (b), (c) and (d) correspond to the sphere size of 16 nm, 8 nm, 4 nm and 2 nm diameter respectively. The upper and lower curves of each image correspond to the sphere placed at position A and B as shown in Fig. 1.

## IV. CONCLUSION

We constructe a simple sphere-tip compression system to systematicly study the local radial properties of ds-DNA. In the system, a model ds-DNA segment of 91 base pairs placed on a rigid solid surface was compressed by a sphere tip gradually falling with a constant velocity. We can obtain a systematic information about ds-DNA by changing the sphere tip position and its size from the diameter 16 nm to 2 nm. The simulation results for the compression sphere of 16 nm diameter are well consistent with the experimental results. The compression properties gradually become sensitive to the sphere-tip size and local DNA conformation position with the decreasing of the sphere tip size. A suddenly force break is shown in the compression-force curve when we decrease the sphere tip size less than or equal to 12 nm diameter. We analyze the folding angle in axis direction and the length between the base pairs in radial direction to understand the local compression property. The analysis shows that a local unwinding process occurs through the coupling of local hydrogen bonding breaking and neighbor base pairs relaxing helix. During the local unwinding process, first the hydrogen bonds between complement base pairs are broken. With the compression aggregating, the local backbones in the compression center are unwound from the double helix conformation to a kind of parallel conformation. This is why the force break appears in the compression-force curves. From these simulation results, we have a clear picture for the mechanically induced conformation transition in radial direction of ds-DNA. This picture provides an alternative method for understanding the local interaction between ds-DNA and protein, and a new method to explore the influence of ds-DNA local property to nanomechanical devices.

This work was supported by National Natural Science Foundation of China under grant No. 11174310, National Basic Research Program of China under grant No. 2012CB932400, and China Postdoctoral Science Foundation Project under grant No. 2012M511159, and the Shanghai Supercomputer Center of

China.